\documentclass{aa}  
\usepackage[table]{xcolor}

\usepackage{graphicx}
\usepackage{txfonts}
\usepackage[colorlinks=True,linkcolor=blue,citecolor=blue]{hyperref}
\usepackage{url}
\usepackage{mathtools}
 \usepackage{booktabs}
\usepackage{multirow}
 \usepackage{placeins}  
\usepackage{siunitx}
\usepackage{etoolbox}
\makeatletter
\patchcmd{\appendix}{\clearpage}{}{}{}
\makeatother
\begin{document}

   \title{Dissecting the Obscured Core of GN20: an Active Galactic Nucleus Outshone by Young Stars}
\author{M. Hamed\inst{1}
\and
L. Colina\inst{1}
\and
P. G. P\'erez-Gonz\'alez\inst{1}
\and
J. \'Alvarez M\'arquez\inst{1}
\and
A. Crespo G\'omez\inst{2}
\and
L.\,A. Boogaard\inst{3}
\and
A. Bik\inst{4}
\and
H. \"Ubler\inst{5}
\and
S. E. I. Bosman\inst{6,7}
\and
A. Alonso-Herrero\inst{8}
\and
M. Perna\inst{1}
\and
S. Arribas\inst{1}
\and
L. Ulivi\inst{1}
\and
M. Annunziatella\inst{9}
\and
L. Costantin\inst{1}
\and
A. Labiano\inst{10}
\and
C.-L. Liao\inst{3}
\and
C. Prieto-Jim\'enez\inst{1,11}
\and
P. van der Werf\inst{3}
\and
F. Walter\inst{7}
}

\institute{Centro de Astrobiolog\'ia (CAB), CSIC-INTA, Ctra. de Ajalvir km 4, Torrej\'on de Ardoz, E-28850, Madrid, Spain
\and
Space Telescope Science Institute (STScI), 3700 San Martin Drive, Baltimore, MD 21218, USA
\and
Leiden Observatory, Leiden University, PO Box 9513, NL-2300 RA Leiden, The Netherlands
\and
Department of Astronomy, Oscar Klein Centre, Stockholm University, AlbaNova University Centre, Stockholm, 106 91, Sweden
\and
Max-Planck-Institut f\"ur extraterrestrische Physik, Gie{\ss}enbachstra{\ss}e 1, D-85748 Garching, Germany
\and
Institute for Theoretical Physics, Heidelberg University, Philosophenweg 12, D-69120 Heidelberg, Germany
\and
Max-Planck-Institut f\"ur Astronomie, K\"onigstuhl 17, D-69117 Heidelberg, Germany
\and
Centro de Astrobiolog\'ia (CAB), CSIC-INTA, Camino Viejo del Castillo s/n, 28692 Villanueva de la Ca\~nada, Madrid, Spain
\and
INAF -- IASF Milano, via A. Corti 12, I-20133 Milano, Italy
\and
Telespazio UK for the European Space Agency (ESA), ESAC, Camino Bajo del Castillo s/n, 28692 Villanueva de la Cañada, Madrid, Spain
\and
Departamento de F\'{i}sica de la Tierra y Astrof\'{i}sica, Facultad de Ciencias F\'{i}sicas, Universidad Complutense de Madrid, E-28040, Madrid, Spain
}

   \date{}

\abstract{}{We investigate the relative contributions of star formation and AGN activity to the total energy budget of GN20, one of the most luminous dusty star-forming galaxies known at $z>4$, through spatially resolved spectral energy distribution decomposition.}{We perform Bayesian SED fitting with CIGALE on two spatially distinct apertures: the nuclear core ($r=0.14\arcsec$, $\sim$1~kpc physical) and the full galaxy ($r=1.4\arcsec$, 9.9~kpc), combining JWST/NIRCam and MIRI broadband imaging, JWST/NIRSpec PRISM IFU pseudo-continuum photometry spanning 42 wavelength bins across rest-frame $0.12$--$1.05~\mu$m, and archival HST and millimeter interferometry data from NOEMA and PdBI.}
{The integrated SED is dominated by stellar-heated dust, with only a marginal AGN contribution at galaxy-wide scales ($f_\mathrm{AGN}^\mathrm{int}=0.09\pm0.02$). The nuclear core, however, requires a significant AGN component ($f_\mathrm{AGN}=0.34\pm0.05$
) to account for a mid-infrared excess at rest-frame $\sim$2.5--3.6~$\mu$m characteristic of AGN-heated torus dust. The AGN accounts for $\sim34\%$ of the nuclear infrared luminosity but only $\sim9\%$ of the total integrated $L_\mathrm{IR}$, explaining its weak signature in integrated diagnostics and its consistency with existing upper limits from \textit{Spitzer} spectroscopy. The inferred black hole mass places GN20 within the local $M_\mathrm{BH}$--$M_\mathrm{bulge}$ relation at the Eddington limit, and in the overmassive regime at sub-Eddington accretion rates, suggesting early and rapid black hole assembly concurrent with the dominant starburst. GN20 exemplifies a class of systems where nuclear-scale SED decomposition, enabled by the angular resolution and infrared sensitivity of JWST, is the only means to uncover a buried AGN overwhelmed by galaxy-wide star formation.}{}

\maketitle
%

\section{Introduction}\label{sec:intro}

Dusty star-forming galaxies (DSFGs) represent one of the most important galaxy populations for understanding cosmic mass assembly and the formation of the most massive systems in the Universe \citep{Chapman2005, Gruppioni2013, Casey2014}. They dominated the dust-obscured cosmic star formation, accounting for more than half of the total star formation rate (SFR) density at $z\sim2$--4, \citep{PG2005, Magnelli2013, Zavala2021}, despite representing only a small fraction of the galaxy population \citep[with space densities of only $\sim$ 10$^{-4}$\,Mpc$^{-3}$,][]{Dudzeviciute2020}. First discovered in submillimeter surveys \citep{Smail1997, Hughes1998, Blain1999}, these systems are characterized by their extreme dust content, which reprocesses ultraviolet (UV) and optical emission from young stars into the far-infrared (FIR) and submillimeter, rendering them faint or invisible at rest-frame UV and optical wavelengths \citep{Casey2014, Gruppioni2020, Hamed21, Barrufet2023, pg2023}.

The peak activity of DSFGs at $z \sim 2$--3 coincides with the peak epoch of supermassive black hole accretion \citep{Richards2006, Hopkins2008, Madau14, Aird2015}, suggesting a fundamental connection between extreme star formation and black hole growth during this critical phase of massive galaxy evolution \citep{Alexander2008, Kormendy2013, Hickox2014}. Understanding the role of AGN in DSFGs is essential for constraining their evolutionary pathways and the mechanisms driving supermassive black hole growth in extreme environments \citep{Casey2014}. However, identifying and characterizing AGN in these heavily obscured systems presents significant observational challenges. These challenges include X-ray absorption by high column densities, dilution of AGN signatures by dominant host star formation, and spatial mixing of nuclear and extended emission in integrated diagnostics.

X-ray observations have revealed AGN activity in a significant fraction of DSFGs, though reported AGN fractions vary considerably depending on the sample, observation depth, and classification criteria \citep{Alexander2005, Pope2006, Laird2010, Wang2013, Franco2018, Stach2019, Uematsu2025}. This variability partly reflects observational challenges inherent to studying heavily obscured systems: X-ray emission can be  absorbed by high column densities \citep{Alexander2005, Daddi2007}, requiring multi-wavelength approaches including mid infrared coverage to identify AGN missed in X-ray surveys \citep{Mateos2012, Lacy2004, Stern2005, Alonso-Herrero2006, Menendez-Delmestre2009, Donley2012}. Many DSFGs are expected to harbor Compton-thick AGN that remain undetected even in deep X-ray observations \citep{Treister2009, Treister2010, Luo2017, DAmato2020}.\\

While the combination of X-ray and infrared diagnostics described above has been successful in identifying AGN candidates, these approaches rely on integrated photometry and therefore face limitations in disentangling nuclear from extended emission in galaxies \citep{DAgostino2019, Barquin-Gonzalez2024}. In spatially extended systems, AGN signatures can be significantly diluted by host galaxy star formation in integrated photometry, which leads to underestimation of AGN contributions \citep{Treister2012}. This spatial mixing particularly affects optical emission line diagnostics \citep{Baldwin1981, Veilleux1987, Kewley2001}, as even moderately luminous AGN can be obscured in integrated line ratios when the host galaxy harbors substantial distributed star formation \citep{Kauffmann2003}. Spatially resolved observations are therefore essential to isolate nuclear emission from surrounding star-forming regions and properly characterize the relative contributions of AGN and star formation across the galaxy \citep{Cresci2015, Venturi2018, Backhaus2023}.\\

The James Webb Space Telescope (JWST) has transformed the study of high-redshift dusty galaxies, providing simultaneous access to rest-frame optical emission lines redshifted into the near-infrared (NIR) and to the rest-frame NIR and mid-infrared (MIR) continuum \citep{Rieke2023, Bohn2023, Boker2022, Arribas2024, Gonzalez-Martin2025}. NIRSpec spatially resolved spectroscopy combined with NIRCam and MIRI imaging now constrains stellar populations, ionized gas, and dust-obscured activity on sub-kpc scales at redshifts beyond the reach of previous facilities \citep{javier23, Colina2023, Marshall2023, Ubler2023, Bik2024, Crespo-Gomez2024}, allowing star formation and AGN activity to be disentangled within individual galaxies.\\

Pre-launch, MIRI was expected to be the instrument best placed to separate obscured nuclear activity from intense star formation in IR-luminous galaxies, accessing MIR diagnostics out to $z\sim4$ that were beyond \textit{Spitzer} for all but the brightest sources \citep{Kirkpatrick2017}. JWST/MIRI surveys have confirmed this expectation, uncovering large populations of obscured AGN out to $z\sim5$, including composite SF--AGN sources whose MIR SEDs cannot be reproduced by star formation alone, with the obscured fraction rising progressively with redshift up to $z\sim4$ \citep{Yang2023, Lyu2024}. Obscured accretion is therefore common among IR-luminous galaxies both locally \citep[e.g.,][]{Alonso-Herrero2012} and at high redshift, but routinely escapes detection in integrated photometry, where the AGN signal is diluted by host emission. JWST/MIRI has also revealed the population of \textit{little red dots} (LRDs): compact, high-$z$ sources with characteristic v-shaped SEDs (blue UV, red rest optical) often showing broad Balmer emission \citep{Labbe2023, Matthee2024, Greene2024, PG2024}. LRDs, and high-$z$ JWST AGN more broadly, frequently host black holes well above the local $M_\mathrm{BH}$--$M_\star$ relation, populating the overmassive regime now widely reported at $z>4$ \citep{Pacucci2023, Harikane2023, Maiolino2024}.\\

The growing evidence for obscured, often overmassive black holes in IR-luminous and compact high-$z$ systems makes it increasingly relevant to ask whether such an AGN component is present, and energetically important, in individual DSFGs. AGN feedback is one of the proposed mechanisms for quenching star formation in massive high-$z$ DSFGs, in scenarios that link them to the compact quiescent galaxies observed at $z\sim2$ \citep{Hopkins2008, Toft2014}. Constraining the AGN content of DSFGs is a necessary step in testing this picture.\\

We apply JWST's spatially resolved spectroscopic and imaging capabilities to one of the most luminous DSFGs known at $z > 4$, GN20 \citep[$z = 4.055$,][]{Pope2005, Daddi2009}, located in the GOODS-North field \citep{Giavalisco2004}. This DSFG has been the subject of extensive multiwavelength observations. Initial submillimeter surveys established GN20 as an exceptionally bright source, with extreme infrared luminosity and a massive dust reservoir \citep[$L_{\rm IR} > 10^{13}$~L$_{\odot}$;][]{Hodge2013, Tan2013, Tan2014, Hodge2015}. High-resolution NOEMA observations revealed complex extended morphology across the molecular gas and dust components \citep{Hodge2013, Boogaard2026}.\\

Recent JWST/MIRI imaging revealed a compact stellar nucleus embedded within an extended stellar envelope, with dust-obscured star formation \citep{Colina2023, Bik2024, Crespo-Gomez2024}. High-resolution ($R \sim 2700$) NIRSpec IFU spectroscopy detected broad H$\alpha$ emission in the nuclear region, which can be interpreted either as a broad-line region of an accreting black hole or as an AGN-driven outflow \citep{Ubler2024}. Furthermore, \citet{Boogaard2026bar} identified a stellar bar-like structure in GN20, which provides a plausible mechanism for funneling gas toward the nucleus and fueling both the central starburst and any AGN activity. SED decomposition, including JWST/MIRI data probing the rest-frame near and MIR, offers an independent and complementary diagnostic to confirm the presence of AGN-heated dust and quantify its energetic contribution, irrespective of the origin of the broad emission line component. However, the energy contribution of AGN relative to star formation remain poorly constrained.\\

In this work, we combine GA-NIFS (ga-nifs.github.io) low-resolution  NIRSpec PRISM ($R \sim 100$) IFU observations, which provide simultaneous coverage of the rest-frame optical continuum and emission lines across the full 0.6--5.3~$\mu$m range, with NIRCam and MIRI imaging and ancillary data to decompose the SED of GN20 into AGN and star-formation components in a nuclear and an integrated aperture, and to quantify the energetic contribution of the AGN relative to the host starburst.\\

The remainder of this paper is structured as follows. Section \ref{sec:data} describes the JWST observations and data reduction procedures, along with the ancillary data. Section \ref{sec:SED} presents our SED technique. We present our results and discuss their implications in Section \ref{sec:results}, and summarize our conclusions in Section \ref{sec:conclusions}. We adopt a flat $\Lambda$CDM cosmology with $\Omega_m = 0.315$ and $H_0 = 67.4$ km s$^{-1}$ Mpc$^{-1}$ \citep{Planck2020}. At the redshift of GN20, this corresponds to a physical scale of $\sim7$ kpc per arcsecond. All stellar masses and SFR assume a \citet{Chabrier2003} initial mass function (IMF) and a stellar population synthesis of \citet{BC03}.

\section{Data}\label{sec:data}

\subsection{Broadband Imaging}\label{sec:broadband}
The multi-wavelength imaging of GN20 used in this work combines archival \textit{HST} data with \textit{JWST} observations and ground-based millimeter interferometry, which covers rest-frame UV through millimeter wavelengths.

\subsubsection{JWST/MIRI and NIRCam Imaging}
GN20 was observed with JWST/MIRI in four filters: F560W, F770W \citep{Colina2023}, F1280W, and F1800W \citep{Crespo-Gomez2024} (program 1264). Observations in F560W, F770W, and F1280W used FASTR1 readout mode with a five-dither medium-size cycling pattern for 1498~s total integration. F1800W was obtained simultaneously during MRS background acquisitions with 1911 s integration using a two-point dither pattern. MIRI imaging data were reduced using the JWST Science Calibration pipeline version 1.12.0 with CRDS context 1140, including additional corrections for striping and background gradients, described in \citet{PG2024} and \citet{ostlin2025}. We refer the reader to \citet{Crespo-Gomez2024} for full details of the reduction procedure for the specific GN20 dataset. Final mosaics have 0.06\arcsec/pixel scale. At $z=4.055$, these filters trace rest-frame $\sim$1-3.8$\mu$m, sampling the stellar continuum and warm dust emission from AGN-heated torus dust.

The \textit{JWST}/NIRCam imaging was obtained in six filters spanning rest-frame $\sim$0.2-$1~\mu$m: F115W, F150W, F200W, F277W, F356W, and F444W \citep{Boogaard2026}. These observations were taken as part of program 1264 (PI: Luis Colina). Imaging in F200W and F356W used 1825~s. total integration distributed over five dithers, while F115W, F150W, F277W, and F444W used 2555~s. over seven dithers. All observations used the SHALLOW4 readout mode with seven groups per integration. NIRCam imaging data were reduced using a custom procedure built on the JWST Science Calibration pipeline version 1.12.3 with CRDS context 1145, including the removal of snowballs and wisps and a background homogenization. We refer the reader to \citet{Boogaard2026bar, Boogaard2026} for full details of the reduction procedure.

\subsubsection{Ancillary Data}
We utilize archival HST/ACS imaging in F435W, F606W, F775W, and F814W, with pixel scale 0.06 \arcsec\ \citep{Hodge2015}. We also included archival NOEMA data at 1.1 mm with synthesized beam 0.26\arcsec\ $\times$ 0.20\arcsec\ \citep{Boogaard2026}, along with the Plateau de Bure Interferometer (PdBI) observation at 880$\mu$m (340 GHz) with synthesized beam 0.30\arcsec\ $\times$ 0.20\arcsec\ at position angle 37$^\circ$ \citep{Carilli2010, Hodge2015}. Both millimeter maps were resampled onto the common $0.06\arcsec$ pixel grid. Each elliptical beam was approximated by a circular Gaussian of FWHM equal to the geometric mean of its major and minor axes, and matched to the F770W PSF by convolution with a circular Gaussian kernel. These millimeter detections constrain the Rayleigh-Jeans tail of the dust SED at rest-frame $\sim$180--220~$\mu$m, complementing the shorter-wavelength FIR coverage described below.

We further include archival MIR and FIR photometry of GN20 from \textit{Spitzer}/MIPS at 24~$\mu$m and \textit{Herschel}/PACS at 100 and 160~$\mu$m and SPIRE at 250, 350, and 500~$\mu$m \citep{Magnelli2013, Tan2014, Liu2018, Crespo-Gomez2024}. These data sample the rest-frame MIR and the peak of the dust thermal emission of GN20. The angular resolution of these data (PACS FWHM $\sim$ 7\arcsec, SPIRE FWHM $\sim$ 18--36\arcsec) is too coarse for nuclear photometric extraction. We therefore use these fluxes as detections in the integrated SED fit ($r=1.4\arcsec$), and as upper limits in the nuclear SED fit ($r=0.14\arcsec$), under the conservative assumption that the nuclear region contributes at most the total galaxy-integrated flux at these wavelengths.
\subsection{NIRSpec IFU Data}\label{sec:nirspec_vs_imaging}

\begin{figure*}[]
\centering
\includegraphics[width=1\textwidth]{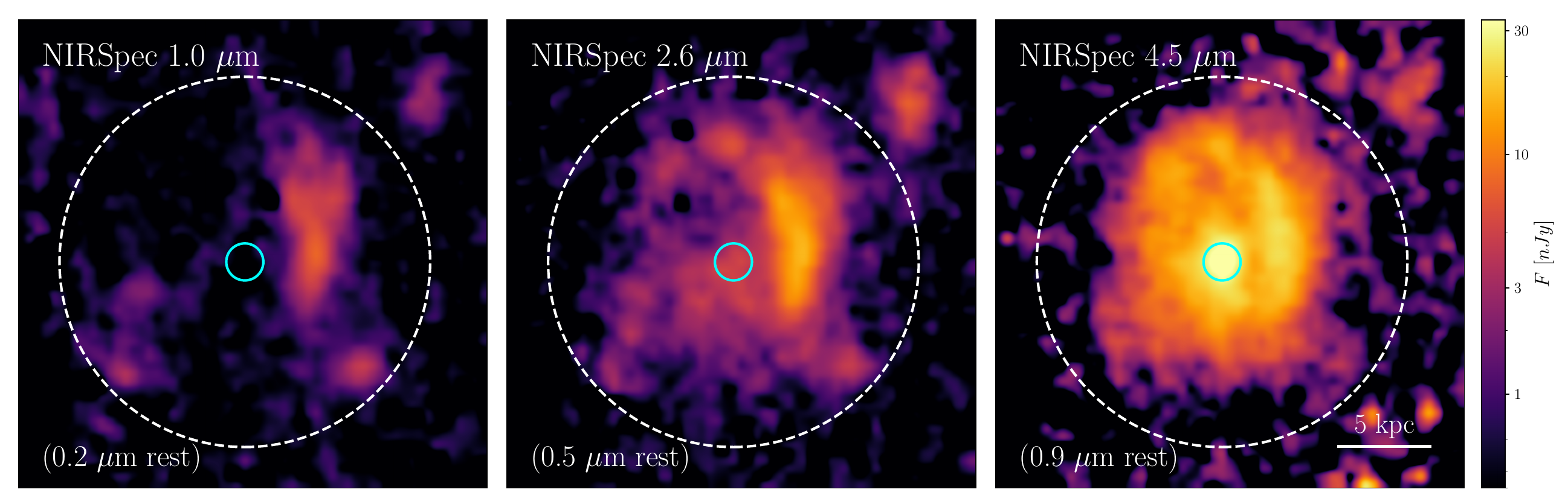}
\caption{JWST/NIRSpec IFU PRISM binned channel maps of GN20 at three representative observed wavelengths sampling the rest-frame UV ($\sim$0.20\,$\mu$m), optical ($\sim$0.51\,$\mu$m), and NIR ($\sim$0.90\,$\mu$m). All maps are shown in units of nJy. The solid cyan circle indicates the nuclear aperture used for core photometry ($r = 0.14\arcsec$), adopted for all bands PSF-matched to the F770W resolution (NIRCam, MIRI F560W, MIRI F770W, and NIRSpec). The dashed white circle marks the integrated photometric aperture ($r = 1.4\arcsec$). A 5\,kpc physical scale bar is shown in the rightmost panel. Rest-frame wavelengths are labeled at the bottom of each panel.}\label{Fig.GN20_stamps}
\end{figure*}

GN20 was observed with JWST/NIRSpec in integral field spectroscopy mode as part of program 1264. Observations were conducted using PRISM/CLEAR with the IRS2RAPID readout mode, providing continuous wavelength coverage from 0.6 to 5.3~$\mu$m with spectral resolution $R \sim 30$--300 \citep{Jakobsen2022}. The observations used a four-point medium cycling dither pattern with one hour total integration time. The 3\arcsec$\times$3\arcsec\ field of view with 0.1\arcsec spatial sampling fully encompasses GN20's spatial extent (Figure \ref{Fig.GN20_stamps}).

At $z = 4.055$, the NIRSpec PRISM observations cover rest-frame $\sim$0.12--1.05~$\mu$m, including critical emission lines: Ly$\alpha$, [O{\sc ii}]$\lambda\lambda$3726,3729, H$\beta$, [O{\sc iii}]$\lambda\lambda$4959,5007, H$\alpha$, and [N{\sc ii}]$\lambda\lambda$6548,6583. Raw data were reduced using the JWST Science Calibration pipeline version 1.11.1 \citep{Ubler2024}.

\begin{table}
\centering
\caption{Key photometric measurements for GN20 integrated aperture extracted with a radius of 1.4\arcsec, and of the GN20 core extracted with the F770W PSF radius of 0.14\arcsec, except for F1280W and F1800W where the radius taken was that of their PSF.}
\label{tab:GN20_photometry}
\begin{tabular}{lcc}
\hline\hline
Instrument/Band & Core  & Integrated \\
&($r=0.14\arcsec$)& ($r=1.4\arcsec$)\\
\hline
\multicolumn{3}{c}{\textit{($\mu$Jy)}} \\
\hline
NIRCam F115W & $-0.010 \pm 0.004$ & $1.073 \pm 0.036$ \\
NIRCam F150W & $-0.016 \pm 0.004$ & $1.494 \pm 0.036$ \\
NIRCam F200W & $-0.034 \pm 0.004$ & $2.468 \pm 0.039$ \\
NIRCam F277W & $0.098 \pm 0.009$ & $4.000 \pm 0.082$ \\
NIRCam F356W & $0.195 \pm 0.009$ & $5.981 \pm 0.140$ \\
NIRCam F444W & $0.375 \pm 0.014$ & $8.072 \pm 0.191$ \\
MIRI F560W & $0.847 \pm 0.028$ & $10.98 \pm 0.303$ \\
MIRI F770W & $1.618 \pm 0.064$ & $18.21 \pm 0.413$ \\
MIRI F1280W & $4.126 \pm 0.457$ & $23.40 \pm 0.700$ \\
MIRI F1800W & $7.813 \pm 0.877$ & $35.36 \pm 3.210$ \\
\hline
\multicolumn{3}{c}{\textit{(mJy)}} \\
\hline
NOEMA 1.1 mm & $1.227 \pm 0.060$ & $10.50 \pm 0.60$ \\
PdBI 880 $\mu$m & $2.016 \pm 0.092$ & $17.40 \pm 1.00$ \\
\hline
\multicolumn{3}{c}{NIRSpec continua (5 of 42 wavelengths, \textit{$\mu$Jy)}} \\
\hline
NIRSpec 1.74 $\mu$m & $-0.013 \pm 0.008$ & $1.425 \pm 0.072$ \\
NIRSpec 2.47 $\mu$m & $0.072 \pm 0.019$ & $3.549 \pm 0.170$ \\
NIRSpec 3.02 $\mu$m & $0.133 \pm 0.013$ & $4.313 \pm 0.120$ \\
NIRSpec 3.72 $\mu$m & $0.278 \pm 0.015$ & $6.369 \pm 0.134$ \\
NIRSpec 5.01 $\mu$m & $0.829 \pm 0.027$ & $12.120 \pm 0.240$ \\
\hline
\multicolumn{3}{c}{Emission line complexes \textit{($\times 10^{-20}$ W\,m$^{-2}$)}} \\
\hline
${\rm H}\alpha+[{\rm N\,{\sc II}}]$ doublet & $0.21 \pm 0.11$ & $7.06 \pm 3.53$ \\
$[{\rm O\,{\sc III}}]$ doublet & $0.04 \pm 0.02$ & $1.84 \pm 0.92$ \\
$[{\rm S\,{\sc II}}]$ doublet & $0.07 \pm 0.04$ & $2.18 \pm 1.09$ \\
\hline
\end{tabular}
\tablefoot{NIRCam, MIRI, and NIRSpec continuum fluxes in $\mu$Jy. Sub-millimeter fluxes (NOEMA, PdBI) in mJy. Negative core fluxes (NIRCam F115W, F150W, F200W, and NIRSpec channels at $\lambda \lesssim 2.2\,\mu$m) indicate non-detections where the measured flux is consistent with the noise. These are treated as $3\sigma$ upper limits in the SED fitting (Section~\ref{sec:SED}). Emission line uncertainties include 50\% systematic uncertainty added in quadrature to account for AGN contamination and spectral blending (see Section~\ref{sec:AGN}). Images were PSF-matched to the PSF of F770W.}\end{table}

\subsection{Data Preparation for SED Fitting}\label{sec:data_prep}

\begin{figure}[]
\centering
\includegraphics[width=0.5\textwidth]{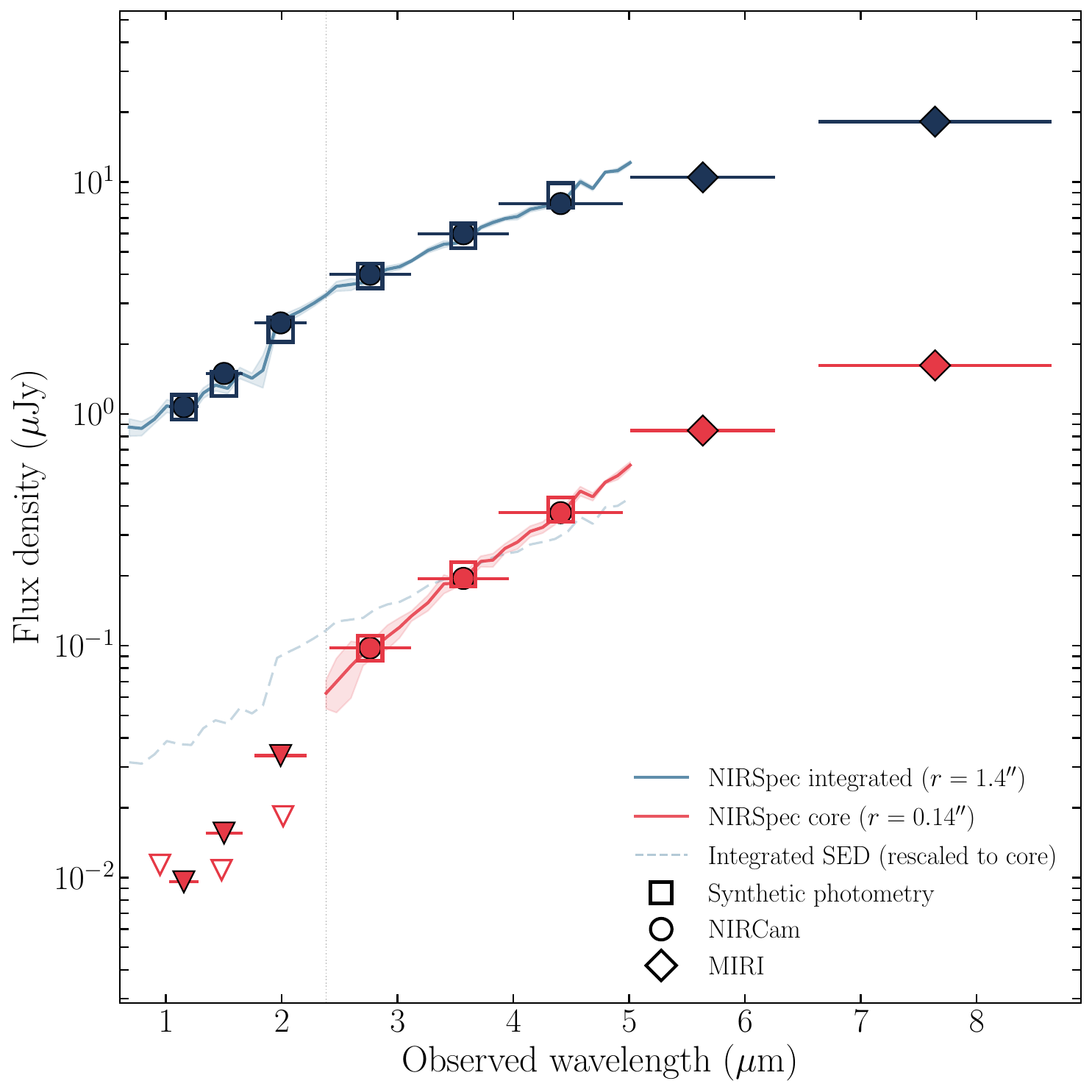}
\caption{NIRSpec continuum fluxes with NIRCam (circles) and MIRI (diamonds) broadband photometry for GN20. The red and blue lines show the NIRSpec pseudo-continuum extracted within the core (r=0.14\arcsec) and integrated (r=1.4\arcsec) apertures, with shaded regions indicating the $\pm1\sigma$ uncertainty. The core spectrum is shown only at wavelengths where it is detected ($\lambda \gtrsim 2.3\,\mu$m), at shorter wavelengths, $3\sigma$ upper limits from binned NIRSpec channels are shown as open downward triangles. NIRCam upper limits are shown as filled downward triangles with horizontal bars indicating the filter bandwidth. Open squares denote synthetic photometry computed by convolving the NIRSpec spectrum with the corresponding filter transmission curves. The dashed blue line shows the integrated SED rescaled to the core flux level to facilitate comparison of the spectral shape between the two apertures. The vertical dotted line marks the wavelength below which the core NIRSpec flux is undetected.}
\label{Fig.NIRSpec_vs_imaging}
\end{figure}
 
To enable spatially resolved SED fitting of GN20's distinct components, all imaging and spectroscopic data were processed to a common spatial resolution and astrometric frame. We verified the astrometry by comparing the position of GN20's compact nuclear emission across all datasets. The NIRCam and MIRI imaging showed consistent centering on the nuclear source. The NIRSpec IFU data cube exhibited a systematic offset of $\sim$-0.21\arcsec\ in RA and $\sim$0.09\arcsec\ in Dec relative to the imaging. We aligned the NIRSpec data cube to the NIRCam F444W imaging by matching the continuum emission peak of the core region at observed $\sim$4.44\,$\mu$m. After applying this correction, the NIRSpec continuum emission at all other wavelengths was verified to be spatially consistent with the corresponding NIRCam bands. All data were resampled to a uniform pixel scale of 0.06\arcsec/pixel.\\

We PSF-matched all maps to the JWST/MIRI F770W resolution (FWHM = 0.28\arcsec, measured from stars in the field), which corresponds to a physical scale of 1.9 kpc at z = 4.055 and enables the separation of GN20's distinct spatial components. Photometry was extracted in two apertures: a nuclear core aperture with r=0.14\arcsec\ ($\sim$1~kpc), corresponding to the F770W PSF half-width at half-maximum (HWHM = FWHM/2), and centered on the brightest F770W pixel \citep{Colina2023}, and an integrated galaxy aperture with r = 1.4\arcsec (9.9 kpc) as in \citet{Crespo-Gomez2024}. For the PSF-matched bands (HST, NIRCam, NIRSpec, MIRI F560W, and F770W), core fluxes were extracted using the r = 0.14\arcsec aperture. MIRI F1280W and F1800W were excluded from the PSF-matched dataset due to their coarser angular resolution (0.42\arcsec\ and 0.59\arcsec\ FWHM respectively). We therefore extract the core photometry of these bands within apertures matched to their native PSF FWHM, assuming that the emission in these bands is dominated by the nuclear point source, and treat the enclosed flux as emission from the compact nuclear component.

To validate this assumption, we compare the F770W radial flux profile (rest-frame $\sim$1.5~$\mu$m) against a point-source PSF model. For a Gaussian PSF with FWHM~$=$~0.28\arcsec, the predicted flux ratio $F(r=0.14\arcsec)/F(r)$ is 63\% at $r=0.21\arcsec$ and 52\% at $r=0.30\arcsec$. The measured F770W ratio at $r=0.21\arcsec$ (the F1280W PSF HWHM) is 63\%, fully consistent with an unresolved nuclear source and confirming that the F1280W core flux contains no significant extended contamination. At $r=0.30\arcsec$ (the F1800W PSF HWHM) the measured ratio is 40\%, suggesting some contribution from surrounding emission at that scale. We note that the F1800W data were obtained with a shallower two-point dither pattern and shorter integration than the other MIRI bands, resulting in a less well-characterized PSF and higher photometric uncertainty at this wavelength. We verify in Section~\ref{sec:results} that treating the F1280W and F1800W fluxes as upper limits does not significantly alter the derived AGN fraction.\\

The NIRSpec PRISM data cube was PSF-matched to the F770W resolution using the wavelength-dependent PSF model from \citet{DEugenio2024}. They characterized the NIRSpec/IFS PRISM PSF using three independent methods: direct stellar observations, serendipitous stars in the IFS field of view, and comparison of synthetic NIRCam images generated from the NIRSpec cube with observed NIRCam imaging. We adopted their empirical wavelength-dependent PSF model, computed the effective circular FWHM at each wavelength, and convolved each spectral slice with a Gaussian kernel to match the F770W PSF. To avoid noise correlation introduced by PSF convolution, photometric uncertainties are derived from the original (non-PSF matched) images \citep{Hamed2026}. For the nuclear core aperture, uncertainties are estimated using the empty aperture method: 2000 apertures of the same radius are placed at random source-free positions on the original image, and the standard deviation of the summed fluxes gives the aperture error directly, naturally accounting for pixel-to-pixel correlations introduced by drizzle resampling. For the integrated aperture, this method is not feasible because the aperture diameter (2.8\arcsec) comprises more than 70\% of the field of view. Instead, we measure the per-pixel RMS in a background annulus of the original image and propagate to the aperture scale.\\

For SED fitting, we separate continuum and emission line contributions to avoid contamination of the stellar continuum. Spectral channels containing emission lines (identified using a mask width of $\pm$15~nm around each line at the observed wavelength) are excluded to construct a pseudo-continuum across the PRISM wavelength range. The continuum-only spectral slices are adaptively binned in wavelength to improve signal-to-noise ratio, with bins spanning 0.1--0.5~$\mu$m and a target S/N of 10. The binning respects wavelength gaps created by emission line removal and does not combine channels across discontinuities. This produces 42 binned pseudo-continuum data points (median bin width = 0.10~$\mu$m), with median S/N of 33.1 for the integrated aperture and 17.9 for the detected core. We use binned pseudo-continuum photometry rather than spectrophotometric SED fitting method, because this approach allows explicit removal of emission line channels prior to fitting, controlled S/N per data point, and avoids resolution-dependent systematics arising from the strongly variable spectral resolution of the PRISM ($R \sim 30-300$) across the wavelength range. Figure \ref{Fig.NIRSpec_vs_imaging} shows the NIRSpec pseudo-continuum binned fluxes along with the NIRCam and MIRI broadband photometry for the core and the integrated apertures.\\

Emission line fluxes were extracted separately from the NIRSpec slices and included as additional constraints in the SED modeling. At the spectral resolution of $R \sim 100$ at 2--4~$\mu$m, several optical emission line complexes are unresolved: H$\alpha$+[N{\sc ii}]$\lambda\lambda$6548,6583, [O{\sc iii}]$\lambda\lambda$4959,5007, and [S{\sc ii}]$\lambda\lambda$6716,6731. For each complex, we selected NIRSpec slices spanning the wavelength range of the line emission, converted each slice from flux density to line flux ($F_\lambda = F_\nu \times c/\lambda^2 \times \Delta\lambda$ with $\Delta\lambda = 50$~\AA), and summed the slices to produce integrated emission line maps. Fluxes were extracted from the same core (0.14\arcsec) and integrated (1.4\arcsec) apertures used for continuum photometry. The blended complexes were input to CIGALE using combined line syntax (e.g., \texttt{line.NII-654.8+H-alpha+NII-658.3}). Summary of the photometric and spectroscopic data is shown in Table \ref{tab:GN20_photometry}.\\

\section{Spectral energy distribution modeling}\label{sec:SED}
\begin{figure*}[]
\centering
\includegraphics[width=1\textwidth]{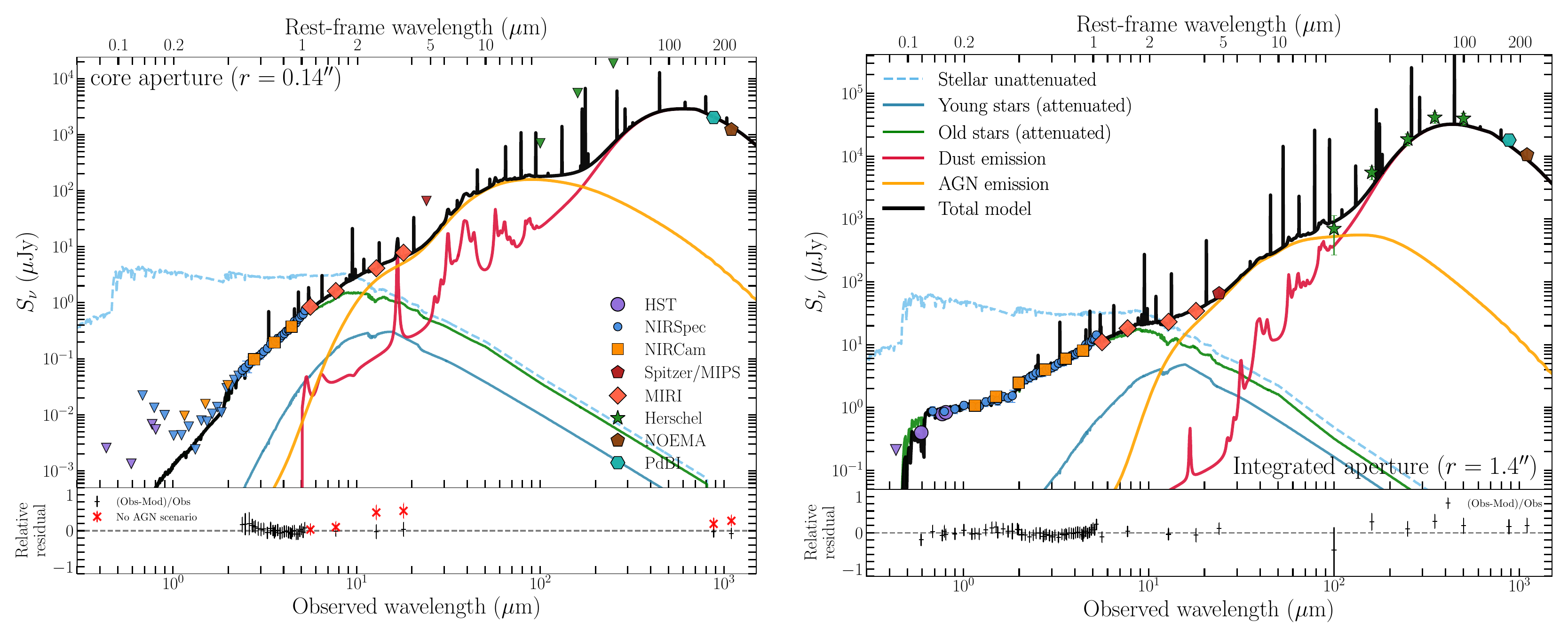}
\caption{Spectral energy distributions of GN20. \textit{Left:} Nuclear core ($r = 0.14^{\arcsec} \sim 1$\,kpc physical at $z = 4.055$). \textit{Right:} Integrated aperture ($r = 1.4^{\arcsec}$, encompassing the entire galaxy). Observed photometry from HST (purple circles), JWST NIRCam (orange squares), JWST MIRI (red diamonds), JWST NIRSpec IFU continuum (blue dots), NOEMA 1.1\,mm (brown pentagon), and PdBI 880\,$\mu$m (teal hexagon). Upper limits shown as downward-pointing triangles with the colors corresponding the observing instrument. Best-fit SED models (black solid) comprise stellar unattenuated (blue), young and old stars attenuated (blue and yellow respectively), dust emission (red), and AGN emission (orange) components. Bottom panels show normalized residuals $(S_{\rm obs} - S_{\rm mod})/S_{\rm obs}$. Reduced $\chi^2$ values: 0.42 (integrated), 0.43 (core). For the core aperture, the no-AGN best-fit model is also shown (red asterisks, residuals only) to illustrate that the MIR excess at F1280W and F1800W cannot be reproduced without an AGN component.}
\label{Fig.SEDs}
\end{figure*}
We perform spectral energy distribution fitting using CIGALE \citep{Boquien2019}, a Bayesian SED fitting code that models galaxy emission from the FUV to the FIR. CIGALE constructs model SEDs by combining stellar population synthesis, nebular emission, dust attenuation, and dust re-emission, based on a defined grid of physical parameters. CIGALE constructs probability distribution functions for the physical parameters using a Bayesian approach \citep{Burgarella2005, Noll2009, Boquien2019}.

In our fitting strategy, we define the separation between young and old stellar populations at 10 Myr, corresponding to the timescale over which massive stars produce ionizing radiation and drive nebular emission.

Nebular continuum and line emission are computed using CLOUDY photoionization models \citep{Ferland2017} that are integrated into CIGALE. We explore ionization parameters log($U$) between $-3$ and $-1$, and set the electron density to $n_e = 100$ cm$^{-3}$. We assume zero Lyman continuum escape fraction ($f_{\rm esc} = 0$) given GN20's high dust content, and set the line width to 300 km s$^{-1}$.\\

We fit GN20's integrated photometry over a 1.4\arcsec\ radius aperture (9.9 kpc at $z = 4.055$), encompassing the full spatial extent of the galaxy. This aperture was also used in \citet{Crespo-Gomez2024}. Additionally, we improve that SED fitting analysis by including the NIRSpec/IFS pseudo-continuum photometry extracted from the line-excluded spectral cube (Section~\ref{sec:data_prep}), providing densely sampled rest-frame optical and NIR constraints on the stellar continuum shape. We adopt the galaxy's spectroscopic redshift $z = 4.055$ \citep{Carilli2011} and apply a 15\% systematic uncertainty added in quadrature to all photometric uncertainties to account for potential calibration systematics and aperture corrections. To investigate the nuclear starburst separately, we also perform SED fitting on a compact nuclear aperture with radius equal to the F770W PSF HWHM (0.14\arcsec\ or 1 kpc), centered on the brightest F770W pixel corresponding to the nuclear center \citep{Colina2023}. This aperture isolates the heavily obscured nuclear starburst from the outer disk and star-forming clumps \citep{Bik2024}. The treatment of the FIR photometry differs between the two apertures. In the integrated fit, the \textit{Spitzer}/MIPS, \textit{Herschel}/PACS, and \textit{Herschel}/SPIRE data are included as detections, providing seven sampling points across the rest-frame 5--100~$\mu$m range that anchor the cold-dust component and the dust luminosity. In the nuclear core fit, the angular resolution of these instruments is insufficient to isolate the central kpc-scale region, and we therefore include the same fluxes as upper limits, encoding the conservative requirement that the nuclear flux cannot exceed the galaxy-integrated value. The NOEMA 1.1~mm and PdBI 880~$\mu$m maps, with $\sim$0.2--0.3\arcsec\ synthesized beams, are used as detections in both apertures.\\

To compute the SEDs with CIGALE, we need to constrain three physical components: the SFH, dust attenuation, and the AGN contribution. The treatment of each is described in subsequent subsections (\ref{sec:SFH}, \ref{sec:dust}, \ref{sec:AGN}).

\subsection{Star Formation History Constraints}\label{sec:SFH}
The rotationally-dominated kinematics of GN20 ($v_{\rm rot}/\sigma_0 \sim 5.3$; \citealt{Ubler2024}), together with the smooth stellar disk morphology revealed by MIRI imaging \citep{Colina2023, Crespo-Gomez2024}, favor sustained star formation over a recent merger-driven starburst. We therefore model the SFH using a delayed exponential as the primary component, with SFR$(t) \propto t \exp(-t/\tau_{\rm main})$, exploring $\tau_{\rm main} = 0.1$--2~Gyr to allow for a range of formation histories from an initial burst to relatively constant star formation over the last Gyr. To test whether a recent starburst component is required by the data, we include an optional recent burst with age $t_{\rm burst} = 1$--10~Myr and mass fraction $f_{\rm burst} = 0.0$--0.99. The inclusion of $f_{\rm burst} = 0.0$ as a parameter allows the SED fitting to determine whether sustained star formation alone can reproduce the observations, or whether an additional recent burst component improves the fit. We verified that even when allowing burst fractions up to 0.99, the best-fit value remains low ($f_{\rm burst} \sim 0.05$), which confirms that the data favor a dominant mature stellar population independently of the prior range adopted.\\

\subsection{Dust Attenuation and Emission Constraints}\label{sec:dust}

GN20 exhibits extreme dust obscuration. \citet{Bik2024} derived $A_V = 14.8 \pm 0.3$ mag from the ratio of dust-corrected Pa$\alpha$-based to integrated infrared-based SFR, indicating that the observed Pa$\alpha$ emission reveals only $11.0 \pm 0.5\%$ of the total SFR. This extreme extinction is primarily associated with GN20's heavily obscured nuclear starburst \citep{Hodge2015, Colina2023, Ubler2024, Boogaard2026bar}.\\

We model dust attenuation of the stellar continuum using the modified \citet{Calzetti2000} attenuation law as described by \citet{Noll2009}, following the approach of \citet{Crespo-Gomez2024}. We explore color excess values $E(B-V)_{\rm young} = 0$--$3$ mag for the young stellar population. This range is appropriate for fitting the flux-weighted average continuum extinction traced by our broadband photometry and pseudo-continuum of NIRSpec, which samples stellar populations distributed across both the nuclear region and the more diffuse outer disk. Following \citet{Noll2009} prescription, we adopt a power-law modification to the attenuation curve with slope $\delta = -0.4$ to $+0.6$ to allow for variations in the wavelength dependence of extinction, and set the UV bump amplitude to zero.\\

We note that star-forming regions may be more deeply embedded in dust than the average stellar population, as indicated by the higher extinction derived specifically from Pa$\alpha$ emission ($A_V = 14.8$ mag; \citealt{Bik2024}). This differential attenuation between nebular and continuum emission is commonly observed in dusty starbursts, where star-forming regions are concentrated in the most obscured nuclear environments while older stellar populations contribute to the less attenuated continuum light \citep{Calzetti2001, Wuyts2013, Reddy2015, Puglisi2017, Shivaei2020}. We account for this differential obscuration using CIGALE's built-in reduction factor parameter in the modified \citet{Calzetti2000} module, which scales the color excess of the older stellar population relative to the young population as $E(B-V)_{\rm old} = f_{\rm att} \times E(B-V)_{\rm young}$, with reduction factor $f_{\rm att} = 0.2$--$1.0$. This parametrization is analogous in effect to the two-component \citet{CharlotFall2000} model but preserves the same attenuation curve shape for both populations, which is better suited to our broadband SED fitting where the curve slope is already a free parameter ($\delta$). 
For the nuclear core aperture, we adopt the unmodified \citet{Calzetti2000} attenuation law, fixing the attenuation slope $\delta = 0$ and the reduction factor to $f_{\rm att} = 0.44$, the canonical value from \citet{Calzetti2000}. The core is undetected at rest-frame UV wavelengths, which prevents meaningful constraints on the attenuation curve slope or the differential obscuration between age components.\\

For dust emission, we employ the \citet{DL2014} models. These are constrained at integrated scales by the broad FIR coverage from \textit{Spitzer}/MIPS, \textit{Herschel}/PACS and SPIRE, NOEMA 1.1~mm, and PdBI 880~$\mu$m, sampling the rest-frame mid-infrared, the peak of the dust SED, and the Rayleigh-Jeans tail. In the nuclear core fit, the same FIR data enter as upper limits given the angular resolution mismatch (Section~\ref{sec:broadband}). We explore minimum radiation field intensities $U_{\rm min} = 0.1$--$50$ to encompass both diffuse and intense heating. The fraction of dust in high radiation field regions ($\gamma$) and the PAH mass fraction ($q_{\rm PAH}$) are well constrained in the integrated fit but less so in the core fit, where the rest-frame $\sim$6--50~$\mu$m regime is bracketed only by upper limits. We adopt ranges motivated by values typical of DSFGs: $\gamma = 0.0$--$0.02$ and $q_{\rm PAH}$ in the range $0.47$--$4.5$ \citep{Magdis2012, Crespo-Gomez2024}, and set the power-law slope of the radiation field distribution to $\alpha = 2.0$.

\subsection{AGN Contribution}\label{sec:AGN}
The presence of a compact and heavily obscured nuclear component in GN20's core \citep{Colina2023, Bik2024, Ubler2024}, with an estimated AGN contribution of $\lesssim$ 15-20\% to the total infrared luminosity \citep{Riechers2014}, suggest a potential buried AGN contribution. To test this possibility, we perform additional SED fitting runs including AGN emission for both the integrated galaxy and the nuclear aperture.\\

We model the AGN torus emission using the \citet{Fritz2006} models, which parameterize the dusty torus geometry and radiative transfer. We adopt a torus outer-to-inner radius ratio $r_{\rm ratio} = 60$, angular distribution parameter $\beta = -0.5$, and radial density gradient $\gamma = 4.0$ following typical values for obscured AGN. We explore optical depths at 9.7~$\mu$m spanning the full \citet{Fritz2006} grid range to allow for moderately to heavily obscured tori. We explore viewing angles $\psi$ ranging from $\ang{0.001}$, to $\ang{89.990}$ between edge-on and face-on AGN. For the polar dust component, we adopt an SMC-like extinction law with $E(B-V)_{\rm polar} = 0.03$ mag.
For the integrated aperture fit, we fix the AGN torus parameters (viewing angle, optical depth, and torus geometry) to the best-fit values derived from the nuclear core fit, allowing only the AGN fraction $f_\mathrm{AGN}$ to vary. This ensures physical consistency between the two apertures, as the torus properties are intrinsic to the central engine and should not change with the extraction aperture.\\

The \citet{Fritz2006} AGN model includes continuum emission from the dusty torus and accretion disk, but does not model emission lines from the AGN. In contrast, CIGALE's \texttt{nebular} module assumes all emission lines arise from star formation. To account for this model limitation and the potential AGN contamination of our observed emission lines, we inflated the uncertainties on all optical emission line fluxes by adding a 50\% systematic uncertainty in quadrature.
This systematic also accounts for uncertainties in extracting line fluxes from blended complexes at $R \sim 100$ spectral resolution, where individual components within each blend cannot be resolved. The inflated uncertainties reduce the weight of emission lines in the fit, which in turn allows the multiwavelength continuum SED to provide the primary constraints on physical parameters. To verify that emission line constraints do not bias the derived physical parameters, we performed an additional fit excluding all emission line fluxes. The resulting best-fit parameters are consistent within the uncertainties with those obtained with the inflated-error approach, confirming that the continuum SED provides the primary constraints on the AGN fraction and stellar population properties.\\

We explore AGN fractions, defined as the fractional contribution of the AGN to the total IR luminosity, with a range of $f_{\rm AGN} = 0.0$--0.9. The inclusion of $f_{\rm AGN} = 0.0$ allows direct comparison between pure starburst and starburst+AGN models. We compare the reduced $\chi^2$ values and examine whether AGN emission significantly improves the fit quality, particularly in the MIR regime where AGN-heated dust emission is most prominent. For the nuclear aperture specifically, we also examine whether a higher AGN fraction is required compared to the integrated galaxy.\\

We note that our photometric coverage in the nuclear core fit extends to rest-frame $\sim$3.6~$\mu$m at the level of detections, with the longer-wavelength FIR data entering only as upper limits. Nevertheless, recent JWST studies have demonstrated robust AGN identification through SED fitting with comparable or shorter rest-frame wavelength baselines: \citet{Delvecchio2025} identified AGN-heated dust at rest-frame 1-3~$\mu$m in stacked Little Red Dots at $z\sim6$ using CIGALE, while \citet{Ronayne2025} found AGN signatures in individual sources at $z\sim5$--9 with MIRI data probing similar rest-frame wavelengths. At rest-frame $\gtrsim$2~$\mu$m, AGN-heated dust produces excess emission over the stellar continuum that can be identified through MIR color selections \citep{Lacy2004, Stern2005} and SED fitting \citep{Leja2018, Ciesla2015}. In our case, the dense sampling (42 NIRSpec pseudo-continuum bins combined with 10 broadband filters) provides strong constraints on the stellar continuum shape, making any MIR excess at F1280W and F1800W highly significant. The independent detection of broad H$\alpha$ emission by \citet{Ubler2024} further corroborates the AGN interpretation.
\section{Results and Discussion}\label{sec:results}
\begin{table}
\centering
\caption{Physical properties of GN20's core, along with the aperture encompassing the full galaxy ($r=1.4^{\arcsec}$), derived from CIGALE SED fitting.}
\label{tab:gn20_properties}
\begin{tabular}{lcc}
\hline\hline
Property & Core  & Integrated  \\
&($r=0.14^{\arcsec}$)& ($r=1.4^{\arcsec}$)\\
\hline
log(M$_\star$ / M$_\odot$) & $10.16 \pm 0.08$ & $11.22 \pm 0.06$ \\
SFR (M$_\odot$/yr) & $163 \pm 47$ & $1987 \pm 187$ \\
log(sSFR / yr$^{-1}$) & $-7.95 \pm 0.20$ & $-7.92 \pm 0.06$ \\
Age$_{\rm mass}$ (Myr) & $243 \pm 116$ & $477 \pm 252$ \\
A(V) (mag) & $4.14 \pm 0.15$ & $2.60 \pm 0.21$ \\
L$_{\text{IR}}$ ($\times 10^{13}$ L$_\odot$) & $0.13 \pm 0.03$ & $1.44 \pm 0.10$ \\
f$_{\rm AGN}$ & $0.34 \pm 0.05$ & $0.09 \pm 0.02$ \\
$\psi_{\rm AGN}$ (deg) & \multicolumn{2}{c}{$14.1 \pm 6.9$} \\
\hline
\end{tabular} 
\end{table}

\subsection{Detection of a dust-buried AGN in the nucleus}
In addition to a nuclear starburst, we explore the presence of an obscured AGN in the bright NIR core of the galaxy. To this end, we perform two CIGALE fits to the nuclear SED: one including an AGN component, whose fraction $f_\mathrm{AGN}$ measures the contribution of the AGN to the total IR luminosity, allowed to vary in the range $f_\mathrm{AGN}=0.0$--$0.9$, and one without an AGN model. The AGN-free model yields a worse fit ($\chi^2_\mathrm{red} = 1.3$), with systematic positive residuals of $\sim$50\% at MIRI F1280W and F1800W (Figure~\ref{Fig.SEDs}, residual of left panel) that cannot be reproduced by any combination of SFH or dust attenuation parameters within the starburst framework. The excess MIR emission at rest-frame $\sim$2.5–3.6 $\mu$m is a characteristic signature of AGN-heated torus dust, and its presence provides direct evidence that a nuclear AGN component is required to explain the core SED.\\

The best-fit SED model (Figure \ref{Fig.SEDs}) including AGN emission yields $0.34 \pm 0.05$ ($\chi^2_\mathrm{red} = 0.43$), with the AGN accounting for roughly one third of the nuclear IR luminosity, $L_\mathrm{IR,core} = (0.13 \pm 0.03)\times 10^{13}\, L_\odot$. The derived nuclear stellar mass is $\log(M_\star/M_\odot) = 10.16 \pm 0.08$, representing $\sim$9\% of the total galaxy stellar mass, while the nuclear SFR of $163 \pm 47\, M_\odot\,\mathrm{yr}^{-1}$ accounts for $\sim$8\% of the integrated star formation. The nuclear dust attenuation reaches $A(V) = 4.14 \pm 0.15$ mag, significantly higher than the integrated value of $2.60$ mag, confirming that the nucleus is the most obscured region of this system. This continuum-derived $A(V)$ is lower than the $A_V = 14.8$~mag derived by \citet{Bik2024} from Pa$\alpha$ emission. As discussed in Section~\ref{sec:dust}, this difference arises because the two tracers probe physically distinct regimes: Pa$\alpha$ is a recombination line emitted within H\,{\sc ii} regions surrounding the youngest, massive stars, which are still deeply embedded in their natal dust clouds, while the continuum-derived $A(V)$ is a luminosity-weighted average over all stellar populations. The ratio $A_V^\mathrm{neb}/A_V^\mathrm{cont} \simeq 3.6$ is larger than the canonical factor of $\sim$2 found in local starbursts \citep{Calzetti2001}, but is not unexpected in a system as extreme as GN20, where the most actively star-forming regions are deeply buried within the nuclear dust concentration while the continuum-derived $A(V)$ is averaged over stellar populations distributed across a range of optical depths, including less obscured regions at larger radii.

The best-fit \citet{Fritz2006} torus parameters favor an obscured viewing geometry ($\psi = 14.1^\circ \pm 6.9^\circ$). 
Independent support for a heavily obscured nuclear environment comes from the X-ray non-detection of GN20 in deep \textit{Chandra} observations, which led \citet{Riechers2014} to infer a Compton-thick column toward the central region. \citet{Riechers2014} estimated the AGN contribution to the total IR luminosity of GN20 at $\lesssim$15-20\%, based on the 6.2~$\mu$m PAH equivalent width measured from integrated photometry \citep{Armus2007}. The apparent tension dissolves when spatial resolution is considered. Our spatially resolved decomposition reveals that the AGN is present in the SED of our smallest aperture for the core of GN20 ($r\sim$1~kpc), where it accounts for $\sim$34\% of the nuclear IR luminosity, but is diluted to $f_\mathrm{AGN}^\mathrm{int} = 0.09 \pm 0.02$ when measured over the full extent of GN20. This is consistent with the $\lesssim$15--20\% upper limit from \citet{Riechers2014}, but provides a direct measurement of the AGN contribution rather than an upper bound, enabled by the dense sampling and spatial resolution of JWST.\\

Independent spectroscopic confirmation of AGN activity is provided by \citet{Ubler2024}, who detected a broad H$\alpha$ component with FWHM$\sim$2000~km~s$^{-1}$ in high-resolution ($R\sim2700$) NIRSpec IFU data, spatially coincident with the nuclear continuum peak. \citet{Ubler2024} consider two interpretations of this feature: emission from a broad-line region surrounding the central black hole, or an AGN-driven outflow propagating above the torus axis. The obscured viewing geometry ($\psi \simeq 14^\circ$) favored by our SED decomposition is consistent with a Type~2 configuration in which the line of sight intercepts the torus, though the broad-line region may be partially visible at this viewing angle. Both interpretations, the BLR emission or AGN-driven outflow, independently confirm the presence of a buried AGN in the nuclear core of GN20. Additional support for AGN ionization comes from the WHAN diagnostic diagram analysis of \citet{Ubler2024}, which reveals ionization consistent with AGN activity extending across large spatial scales in GN20.\\

The AGN signature is detected within our smallest aperture ($r=0.14\arcsec \sim$1~kpc). In the integrated SED, which is dominated by star-formation heated dust distributed across the full $\sim$10~kpc extent of the galaxy, the AGN signal is heavily diluted. This is further compounded by the 15\% systematic uncertainty added in quadrature to all integrated photometric fluxes, which renders a $\sim$9\% AGN contribution in the integrated aperture.

As a further test of the robustness of the AGN detection, we repeated the nuclear CIGALE fit treating the F1280W and F1800W fluxes as upper limits rather than detections, to account for any potential contribution from extended emission at the coarser angular resolution of these bands (Section~\ref{sec:data_prep}). The resulting AGN fraction ($f_\mathrm{AGN} = 0.29 \pm 0.10$) is consistent with the fiducial value within the uncertainties, confirming that the AGN detection is robust against the treatment of the longer-wavelength MIRI photometry.
\subsection{AGN bolometric luminosity and black hole mass}

The nuclear AGN fraction of $f_\mathrm{AGN} = 0.34 \pm 0.05$ implies an AGN bolometric luminosity of
\begin{align}
L_\mathrm{AGN} &= f_\mathrm{AGN} \times L_\mathrm{IR,core} = (4.4 \pm 1.2)\times 10^{11}\, L_\odot \label{eq:Lagn}\\
&\simeq (1.7 \pm 0.5)\times 10^{45}\, \mathrm{erg\,s^{-1}}. \nonumber
\end{align}
Assuming the black hole is accreting at the Eddington limit ($\lambda_\mathrm{Edd} = 1$), we obtain a black hole mass of
\begin{equation}
M_\mathrm{BH} \simeq \frac{L_\mathrm{AGN}}{1.26\times10^{38}\,\mathrm{erg\,s^{-1}}} \simeq 1.3 \times 10^{7}\, M_\odot, \label{eq:Mbh}
\end{equation}
or $\log(M_\mathrm{BH}/M_\odot) \simeq 7.1$.\\

The best-fit \citet{Fritz2006} torus parameters yield a viewing angle of $\psi = 14.1^\circ \pm 6.9^\circ$ (Table~\ref{tab:gn20_properties}), consistent with an obscured AGN viewed through the torus. \citet{Ubler2024} detected a broad H$\alpha$ component with FWHM$\sim$2000~km~s$^{-1}$ in the nuclear region, which they interpret as either BLR emission or an AGN-driven outflow. Under the BLR scenario, their virial estimate of $\log(M_\mathrm{BH}/M_\odot) = 7.3 \pm 0.4$ is consistent with our Eddington-limit value. Our SED-based AGN detection is independent of the origin of the broad line.\\

Comparing this black hole mass to the nuclear stellar mass ($\log(M_\star^\mathrm{core}/M_\odot) = 10.16$) yields $M_\mathrm{BH}/M_\star \simeq 0.09\%$, placing GN20 within the scatter of the local $M_\mathrm{BH}$--$M_\mathrm{bulge}$ relation \citep{Kormendy2013}. We note that the local relation is defined for bulge stellar masses, and our nuclear core stellar mass measured within $r \sim 1$~kpc serves as a proxy at this redshift.\\

\citet{Riechers2014} estimated $\log(M_\mathrm{BH}/M_\odot) = 8.1$--$8.5$ also at the Eddington limit, derived from the rest-frame 6~$\mu$m AGN continuum luminosity of the integrated \textit{Spitzer}/IRS spectrum via the local $\nu L_\nu(6\,\mu\mathrm{m})$--$L_\mathrm{X}$(2--10~keV) relation and a bolometric correction $\kappa_{2\text{-}10\,\mathrm{keV}} = 55$. Their Eddington-limit mass is an order of magnitude higher than our estimate, a discrepancy that likely reflects both the use of galaxy-integrated photometry, which overestimates the AGN luminosity by folding in starburst-heated warm dust (Section~\ref{sec:integrated_SED}), and the additional uncertainties inherent to the conversion from MIR luminosity to black hole mass. Our spatially resolved nuclear SED decomposition provides a more direct constraint on the AGN luminosity, and we therefore consider $\log(M_\mathrm{BH}/M_\odot) \simeq 7.1$ the more robust black hole mass.

Taken together with the evidence for a compact nuclear stellar component, a stellar bar-like structure \citep{Boogaard2026bar}, and an extended dusty rotating disc actively forming stars at large radii \citep{Hodge2013, Colina2023, Ubler2024, Bik2024}, GN20 hosts a structured, multi-component system with an actively growing central black hole already in place at $z \sim 4$.\\

\subsection{Integrated SED: An AGN outshone by an extended starburst}\label{sec:integrated_SED}
Having established the presence of a buried AGN in the nuclear core, we now examine the integrated SED of GN20 to assess whether any AGN signature is detectable on galaxy-wide scales. We perform SED fitting of the integrated aperture (r=1.4$\arcsec$, encompassing 9.9 kpc) both with and without an AGN component (Section \ref{sec:AGN}). With the AGN component allowed to vary in the models, the fit converges to a small AGN contribution of $f_\mathrm{AGN}^\mathrm{int} = 0.09 \pm 0.02$. The corresponding integrated AGN bolometric luminosity is $L_\mathrm{AGN}^\mathrm{int} \simeq (1.3 \pm 0.4)\times 10^{12}\, L_\odot$. The AGN contribution to the total integrated infrared luminosity is therefore at the few-percent level, well within the upper limit of $\lesssim 15$--$20\%$ derived by \citet{Riechers2014} from \textit{Spitzer}/IRS spectroscopy. \citet{Crespo-Gomez2024} also reported no significant AGN contribution in the integrated imaging fit of GN20 using \texttt{CIGALE}. Our analysis extends their work by incorporating JWST/NIRSpec PRISM IFU pseudo-continuum photometry (42 wavelength bins spanning rest-frame $0.12$--$1.05~\mu$m) and NOEMA 1.1~mm and PdBI 880~$\mu$m continuum data, which provide denser sampling of the rest-frame optical and NIR continuum and enable the marginal recovery of the otherwise heavily diluted AGN signal. The best-fit SED model for the integrated aperture yields a total stellar mass of $\log(M_\star/M_\odot) = 11.22 \pm 0.06$ and a SFR of $1987 \pm 187\, M_\odot\,\mathrm{yr}^{-1}$ (Table~\ref{tab:gn20_properties}).\\

The integrated SFR is in agreement with the IR-based measurement of $\mathrm{SFR}_\mathrm{IR} = 1860 \pm 90\, M_\odot\,\mathrm{yr}^{-1}$ \citep{Tan2014}. The observed (uncorrected) Pa$\alpha$-based star formation rate of $\mathrm{SFR}_\mathrm{Pa\alpha}^\mathrm{obs} \sim 205\, M_\odot\,\mathrm{yr}^{-1}$ \citep{Bik2024, Bik2025corr} represents $\sim$11\% of the integrated value, consistent with the extreme dust obscuration of this system. Our SED fitting recovers $L_\mathrm{IR} = (1.44 \pm 0.10)\times 10^{13}\, L_\odot$, consistent within uncertainties with the $(1.9 \pm 0.4)\times 10^{13}\, L_\odot$ derived by \citet{Riechers2014}, and with the $(1.86 \pm 0.09)\times 10^{13}\, L_\odot$ from \citet{Tan2014}. The derived stellar mass is consistent with previous estimates for GN20 \citep[$\log(M_\star/M_\odot) = 10.93 \pm 0.30$ and $11.04 \pm 0.20$, respectively]{Crespo-Gomez2024, Tan2014}. We find $A(V) = 2.60 \pm 0.21$~mag, higher than the $1.50 \pm 0.01$~mag reported by \citet{Crespo-Gomez2024} with \texttt{CIGALE} without NIRSpec constraints. The denser sampling of the rest-frame optical and NIR continuum provided by the 42 NIRSpec IFU pseudo-continuum points tightens the constraint on $A(V)$, which directly affects the stellar mass estimate. Our results are also in broad agreement with the earlier estimate of $\log(M_\star/M_\odot) = 11.36 \pm 0.20$ by \citet{Daddi2009} based on eight photometric bands spanning HST/ACS to \textit{Spitzer}/IRAC.\\

Our SED model predicts a continuum flux density of $13.5\pm2.0$~mJy at rest-frame 205~$\mu$m (observed $\sim$1.04~mm), in agreement within $\sim$1$\sigma$ with the $16.2 \pm 2.5$~mJy measured by \citet{Kolupuri2025} from [NII]~205~$\mu$m continuum observations with the IRAM interferometer using a larger aperture (beam of 1.78$\arcsec\times$1.65$\arcsec$). We get [N{\sc ii}]~205~$\mu$m line flux of $(2.08\pm0.94)\times10^{-20}$~W~m$^{-2}$, in excellent agreement with the $(2.22\pm0.68)\times10^{-20}$~W~m$^{-2}$ derived from the line flux of $2.3\pm0.7$~Jy~km~s$^{-1}$ measured by \citet{Kolupuri2025}.\\

\citet{Crespo-Gomez2024} performed integrated SED fitting of GN20 including \textit{Spitzer}/MIPS and \textit{Herschel}/PACS and SPIRE photometry spanning rest-frame $\sim$5--120~$\mu$m, with no AGN model at galaxy-wide scales. The denser sampling of the rest-frame optical and NIR continuum from our NIRSpec data, combined with the same FIR constraints (used as detections at integrated scales and as upper limits at the nuclear scale, Section~\ref{sec:broadband}), enables the marginal recovery of an otherwise heavily diluted AGN signal at the few-percent level.\\

\subsection{GN20 in the context of obscured AGN at high redshift}
GN20 adds to the growing population of obscured AGN in massive, IR-luminous galaxies whose nuclear AGN signatures are heavily diluted in galaxy-integrated diagnostics. Pre-JWST studies of IR-bright galaxies at $z<3$ found that more than $40\%$ of IR-selected samples host an AGN, with composite systems where both star formation and AGN contribute to the MIR comprising $\sim$30\% of the population \citep{Kirkpatrick2015}. Similarly, \citet{Riechers2014} found AGN contributions of $\sim$11--36\% to $L_\mathrm{IR}$ in $z\sim2$ SMGs with comparable mid-infrared AGN fractions, broadly consistent with GN20's integrated value.\\
JWST/MIRI has now extended this census to higher redshifts and lower luminosities. Using the eight-band SMILES survey, \citet{Lyu2024} identified a large population of AGN candidates in the GOODS-S/HUDF field, with the obscured AGN fraction increasing progressively from $z\sim0$ to $z\sim4$ \citep{Lyu2024, Alberts2024}. These results indicate that a fraction of accreting supermassive black holes coexist with intense starbursts across cosmic time, with their signatures diluted in galaxy-wide photometry.\\

The GN20 fit illustrates how integrated photometry can mask a nuclear AGN. As shown in Section~\ref{sec:integrated_SED}, our integrated $r=1.4\arcsec$ fit converges to $f_\mathrm{AGN}^\mathrm{int} \approx 0.09$ that is much smaller than the AGN contribution recovered at the nuclear scale. The MIR excess driving the AGN identification at $r=0.14\arcsec$ is diluted by a factor $\sim$4 in the integrated aperture by the surrounding star-forming and stellar mass regions, which dominate the rest-frame NIR to FIR emission of the system. Spatially resolved SED decomposition is therefore necessary to identify buried AGN in DSFGs.

The AGN bolometric luminosity of $L_\mathrm{AGN} \simeq 1.7 \times 10^{45}\,\mathrm{erg\,s^{-1}}$ would be detected in galaxies with typical SFR at similar redshifts \citep{Madau14}. The AGN is outshone in GN20 because the host starburst ($L_\mathrm{IR} \sim 1.4 \times 10^{13}\, L_\odot$) overwhelms the nuclear emission. This suggests that AGN of comparable luminosity hosted by DSFGs may be systematically missed in precisely the systems where black hole growth is more active.\\

\begin{figure}[]
\centering
\includegraphics[width=0.5\textwidth]{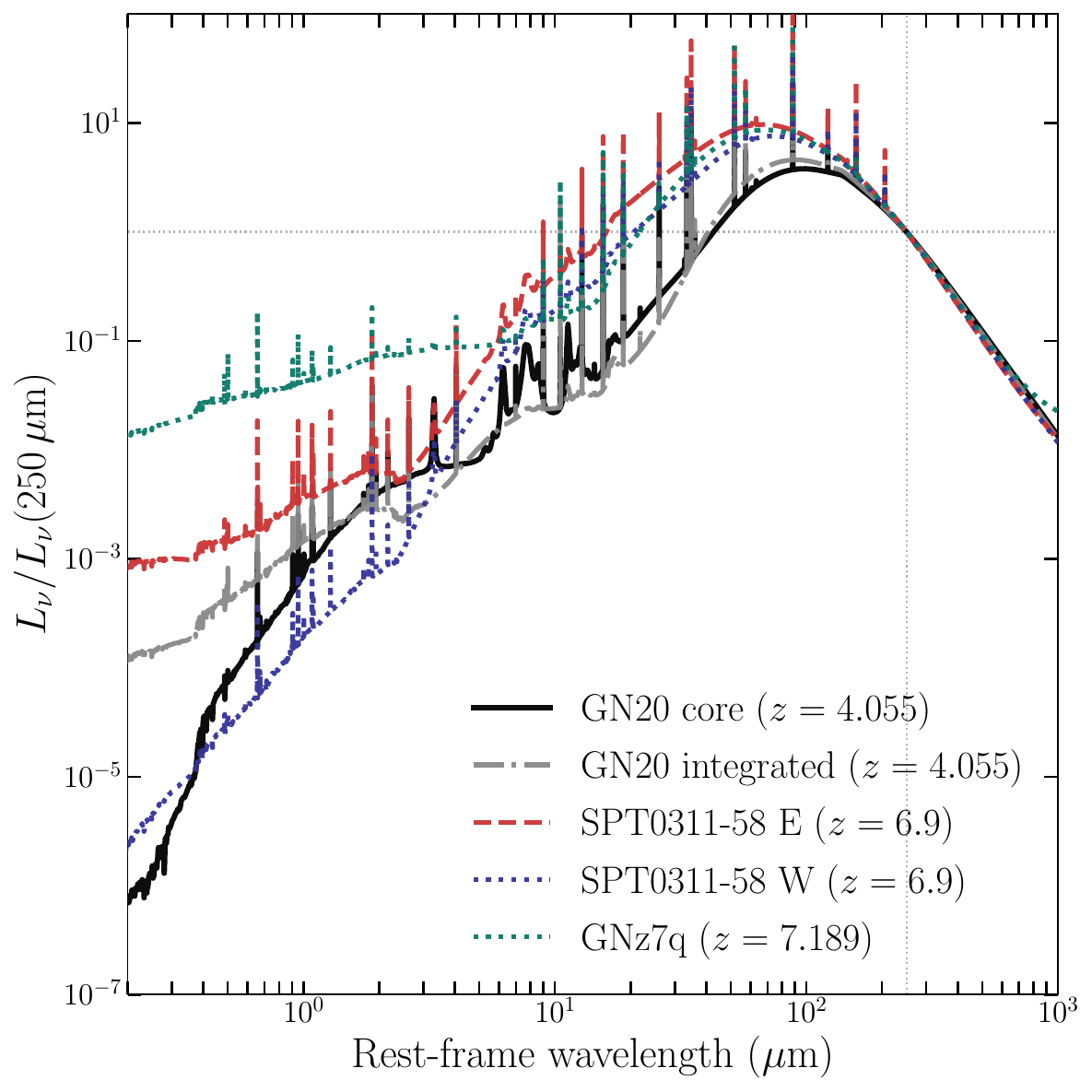}
\caption{Rest-frame SED comparison between GN20, SPT0311-58 \citep{javier23}, and GNz7q \citep{Fujimoto2022, Fei2026}. CIGALE best-fit models for the GN20 core (black solid) and integrated aperture (grey dash-dotted) at $z=4.055$ are shown alongside SPT0311-58 E (red dashed) and W (blue dotted) at $z=6.9$, and GNz7q (green) at $z=7.189$. All SEDs are normalized at rest-frame $250\,\mu$m, within the cold-dust Rayleigh-Jeans tail where AGN emission is negligible. Both SPT0311-58 galaxies and GNz7q show a MIR excess relative to their cold-dust emission, comparable to the excess that requires an AGN component in our GN20 core fit. GNz7q, recently confirmed as a super-Eddington red quasar hosted by a massive starburst \citep{Fei2026}, exhibits the most pronounced MIR excess, consistent with its dominant AGN contribution.}\label{Fig.comparison}
\end{figure}

Figure~\ref{Fig.comparison} compares the GN20 SEDs (the core and the integrated) with the two components (E and W) of SPT0311-58 \citep{javier23} at $z=6.9$ and with GNz7q \citep{Fujimoto2022, Fei2026} at $z=7.189$. After normalizing at rest-frame 250 $\mu$m, where AGN dust emission is negligible, all of these SEDs except the GN20 integrated aperture show a MIR excess relative to the cold-dust component. In GN20 the excess is precisely what is diluted beyond recognition at galaxy-wide scales. GNz7q has recently been spectroscopically confirmed with JWST as a super-Eddington accreting black hole ($\lambda_\mathrm{Edd} = 2.7$, $\log(M_\mathrm{BH}/M_\odot) = 7.55$) embedded in a massive starburst host with $\mathrm{SFR} \sim 330\, M_\odot\,\mathrm{yr}^{-1}$ \citep{Fei2026}. In both GN20's nuclear core and GNz7q, the AGN contributes significantly to the MIR, but in GN20 this excess is detectable only in the spatially resolved core aperture and is diluted beyond recognition at galaxy-integrated scales. SPT0311-58 shows a MIR excess comparable in shape to GN20's nuclear SED, despite having been modeled with pure-starburst components only \citep{javier23}. An AGN-inclusive fit was not performed for SPT0311-58, and the question of a buried AGN in that system remains open.

GN20 illustrates a broader methodological point. Obscured AGN fractions inferred from JWST surveys at $z\gtrsim2$ \citep{Lyu2024, Alberts2024} that rely on galaxy-integrated SED fitting may be conservative, since a buried AGN contribution can be partially accommodated by warm-dust starburst components when AGN models are not included. The most luminous, dust-rich, and morphologically complex high-redshift systems, in which rapidly accreting central black holes might be present, are those for which integrated SED diagnostics are least sensitive to a nuclear AGN. Spatially-resolved SED decomposition with AGN model components, applied here to the nuclear core of GN20, is required to identify and characterize this hidden population.
\section{Conclusions}\label{sec:conclusions}
We have presented a spatially resolved SED decomposition of the $z=4.055$ dusty star-forming galaxy GN20, combining JWST/NIRCam, MIRI, and NIRSpec PRISM IFU observations with archival HST and millimeter interferometry data. Using CIGALE Bayesian SED fitting across two apertures, the nuclear core ($r=0.14\arcsec$, $\sim$1~kpc) and the full galaxy ($r=1.4\arcsec$, 9.9~kpc), we characterize the relative contributions of star formation and AGN activity to the total energy budget of GN20.\\

The integrated SED is dominated by stellar-heated dust, with a marginal AGN signature at galaxy-wide scales ($f_\mathrm{AGN}^\mathrm{int} = 0.09 \pm 0.02$). The best-fit model yields a total stellar mass of $\log(M_\star/M_\odot) = 11.22 \pm 0.06$, a SFR of $1987 \pm 187\, M_\odot\,\mathrm{yr}^{-1}$, and $L_\mathrm{IR} = (1.44 \pm 0.10)\times 10^{13}\, L_\odot$, in good agreement with previous estimates \citep{Tan2014, Bik2024, Bik2025corr}. In contrast, the nuclear core SED requires a significant AGN component ($f_\mathrm{AGN} = 0.34 \pm 0.05$) to reproduce the observed MIR excess at rest-frame $\sim$2.5--3.6~$\mu$m.\\

The nuclear SFR of $163 \pm 47\, M_\odot\,\mathrm{yr}^{-1}$ within the central $\sim$1~kpc implies a nuclear SFR surface density of $\Sigma_\mathrm{SFR}^\mathrm{core} \sim 52\, M_\odot\,\mathrm{yr}^{-1}\,\mathrm{kpc}^{-2}$, approximately 8 times higher than the galaxy-wide average of $\sim$6.5~$M_\odot\,\mathrm{yr}^{-1}\,\mathrm{kpc}^{-2}$ estimated over the full 9.9~kpc-radius aperture, consistent with the compact, heavily obscured nuclear starburst identified through MIRI imaging \citep{Colina2023}.\\

The AGN contributes $\sim$34\% of the nuclear $L_\mathrm{IR}$ ($L_\mathrm{AGN} \simeq 4.4 \times 10^{11}\, L_\odot$), and $\sim$9\% of the total integrated $L_\mathrm{IR}$, which is dominated by star-formation heated dust distributed across the full $\sim$10~kpc extent of GN20. This spatial dilution, compounded by the 15\% systematic photometric uncertainties applied to the integrated SED fitting, explains why the AGN is only marginally detectable in integrated diagnostics and is consistent with the $\lesssim$15--20\% upper limit on the AGN contribution to the total $L_\mathrm{IR}$ derived by \citet{Riechers2014} from \textit{Spitzer}/IRS spectroscopy.\\

The AGN bolometric luminosity implies a black hole mass of $\log(M_\mathrm{BH}/M_\odot) \simeq 7.1$ at the Eddington limit, consistent within the uncertainties with the virial estimate of $\log(M_\mathrm{BH}/M_\odot) = 7.3 \pm 0.4$ derived by \citet{Ubler2024}, itself a lower limit given uncorrected BLR extinction. Using the nuclear core as proxy for bulge mass, $M_\mathrm{BH}/M_\star \simeq 0.09\%$, consistent with the local scaling relation \citep{Kormendy2013}. At $\lambda_{\rm Edd} \sim 0.1$, these rise to $\log(M_{\rm BH}/M_\odot) \sim 8.1$ and $M_{\rm BH}/M_\star \sim 0.9\%$, placing GN20 in the overmassive regime at $z > 4$, suggesting that the central black hole assembled earlier and more rapidly than its host stellar mass.\\

GN20 represents a compelling case where a moderately luminous AGN is largely hidden from integrated diagnostics by the overwhelmingly dominant global star formation. Spatially resolved SED decomposition, enabled by the angular resolution and wavelength coverage of JWST, is essential to uncover such buried AGN in extreme high-redshift starbursts, and underscores the importance of nuclear-scale analysis in characterizing the energetics and evolutionary pathways of the most massive galaxy progenitors in the early Universe.\\

\begin{acknowledgements}
M.H., L.C., and J.Á.-M. acknowledge support from grant PIB2021-127718NB-100, P.G.P.-G. acknowledges support from grant PID2022-139567NB-I00, and J.Á.-M. acknowledges support from grant PID2024-158856NA-I00 funded by Spanish Ministerio de Ciencia e Innovaci\'on MCIN/AEI/10.13039/501100011033, FEDER {\it Una manera de hacer Europa}. This work has made use of the Rainbow Cosmological Surveys Database, which is operated by the Centro de Astrobiología (CAB), CSIC-INTA. AAH acknowledges support from grant PID2021-124665NB-I00, funded by MCIN/AEI/10.13039/501100011033 and by ERDF A way of making Europe. The project that gave rise to these results received the support of a fellowship from the “la Caixa” Foundation (ID 100010434). The fellowship code is LCF/BQ/PR24/12050015. SEIB is supported by the Deutsche Forschungsgemeinschaft (DFG) under Emmy Noether grant number BO 5771/1-1a. L.A.B. acknowledges support from the Dutch Research Council (NWO) under grant VI.Veni.242.055 (\url{https://doi.org/10.61686/LAJVP77714}).  H\"U acknowledges support by the Max Planck Society through the Lise Meitner Excellence Program. H\"U acknowledges funding by the European Union (ERC APEX, 101164796). Views and opinions expressed are however those of the authors only and do not necessarily reflect those of the European Union or the European Research Council Executive Agency. Neither the European Union nor the granting authority can be held responsible for them. MP acknowledges support through the grants PID2021-127718NB-I00, PID2024-159902NA-I00, and RYC2023-044853-I, funded by the Spain Ministry of Science and Innovation/State Agency of Research MCIN/AEI/10.13039/501100011033 and El Fondo Social Europeo Plus FSE+. A.B.,  acknowledges support from the Swedish National Space Administration (SNSA).
\end{acknowledgements}

\bibliographystyle{aa}
\bibliography{aanda.bib}

\end{document}